\documentclass[%
superscriptaddress,
preprint,
nofootinbib,
 amsmath,amssymb,
 aps,
prd,
]{revtex4}

\usepackage{graphicx}
\usepackage{dcolumn}
\usepackage{bm}
\usepackage{hyperref}
\usepackage{physics}
\usepackage{breqn}
\usepackage{tabularx}

\usepackage{color}

\usepackage{cancel}
\usepackage{ulem}

\usepackage{comment}

\begin{document}

\title{Lattice QCD studies on decuplet baryons as meson-baryon bound states in the HAL QCD method}

\newcommand{\Kyoto}{
Center for Gravitational Physics and Quantum Information, 
Yukawa Institute for Theoretical Physics, Kyoto University, Kitashirakawa Oiwakecho, Sakyo-ku, 
Kyoto 606-8502, Japan}

\newcommand{\Riken}{RIKEN Nishina Center, Wako 351-0198, Japan}

\newcommand{\iTHEMS}{Interdisciplinary Theoretical and Mathematical Sciences Program (iTHEMS), RIKEN, Wako 351-0198, Japan}

\newcommand{\Cider}{Division of Scientific Information and Public Policy, Center for Infectious Disease Education and Research, Osaka University, 2-8 Yamadaoka, Suita City, Osaka 565-0871, Japan}

\author{Kotaro~Murakami}
 \affiliation{\Kyoto}
 \affiliation{\iTHEMS}
\author{Yutaro~Akahoshi\footnote{Present address: Quantum Laboratory, Fujitsu Research, Fujitsu Limited. Kanagawa 211-8588, Japan}}
   \affiliation{\Kyoto}
\author{Sinya~Aoki}
   \affiliation{\Kyoto}
\author{Takumi~Doi}
    \affiliation{\iTHEMS}
\author{Kenji~Sasaki}
   \affiliation{\Cider}

\collaboration{HAL QCD Collaboration}

\date{\today}

\begin{abstract}
  We study decuplet baryons from meson-baryon interactions in lattice QCD, in particular, $\Delta$ and $\Omega$ baryons from P-wave $I=3/2$ $N\pi$ and $I=0$ $\Xi\bar{K}$ interactions, respectively. Interaction potentials are calculated in the HAL QCD method using 3-quark-type source operators at $m_{\pi} \approx 410~\textrm{MeV}$ and $m_{K} \approx 635~\textrm{MeV}$, where $\Delta$ as well as $\Omega$ baryons are stable. We use the conventional stochastic estimate of all-to-all propagators combined with the all-mode averaging to reduce statistical fluctuations. We have found that the $\Xi\bar K$ system has a weaker attraction than the $N\pi$ system while the binding energy from the threshold is larger for $\Omega$ than $\Delta$. This suggests that an inequality $m_{N}+m_{\pi}-m_{\Delta}<m_{\Xi}+m_{\bar K}-m_{\Omega}$ comes mainly from a smaller spatial size of a $\Xi \bar K$ bound state due to a larger reduced mass, rather than its interaction. Root-mean-square distances of bound states in both systems are small, indicating that $\Delta$ and $\Omega$ are tightly bound states and thus can be regarded qualitatively as composite states of 3 quarks.
Results of binding energies agree with those obtained from temporal 2-point functions within large systematic errors, which arise dominantly from the lattice artifact at short distances.
\end{abstract}

                   
\preprint{YITP-22-113, RIKEN-iTHEMS-Report-22}
\maketitle


\section{Introduction}
Most of hadrons have been understood well in the quark model, while exceptions, called exotic hadrons, were found in various experiments recently~\cite{Brambilla:2019esw}. Exploring properties or internal structures of such exotic hadrons from QCD is one of the biggest issues in hadron physics. Since exotic hadrons typically appear as resonances due to non-perturbative QCD interactions, theoretical studies from first-principles lattice QCD are mandatory.

Properties of hadron resonances such as masses and decay rates are evaluated from hadron scatterings. The conventional method to analyze hadron scatterings in lattice QCD is the finite-volume method~\cite{Luscher:1990ux, Rummukainen:1995vs}, which relates energies on finite volume(s) to scattering amplitudes on the infinite volume. An alternative method is the HAL QCD method~\cite{Ishii:2006ec, Aoki:2009ji, Ishii:2012ssm}, in which we extract interaction potentials directly in lattice QCD and then obtain scattering amplitudes from potentials by solving the Schor\"{o}dinger equations in the infinite volume. This method is particularly advantageous to study systems involving baryons~\cite{Iritani:2018vfn}.

As a first step toward understandings of exotic hadrons including pentaquarks, we focus our attention on decuplet baryons, spin $3/2$  baryons symmetric under quark flavor exchanges, since all decuplet baryons except $\Omega$ appear as resonances. There are studies on decay properties of the process $\Delta \to N\pi$ by extracting the transfer 
matrix elements in lattice QCD~\cite{Alexandrou:2013ata, Alexandrou:2015hxa}, and are also several studies on the $N\pi$ scatterings for the $\Delta$ baryon in the finite-volume method  at the lighter quark masses~\cite{Andersen:2017una, Paul:2018yev, Andersen:2019ktw, Silvi:2021uya, Bulava:2022vpq}, where
signals of the $\Delta$ as a resonance are observed.

In this paper, as a first analysis for decuplet baryons in the HAL QCD method,
we investigate a question why $\Omega$ appears as a stable particle below the $\Xi\bar{K}$ threshold while others such as $\Delta$ become resonances. 
Historically, 
the decuplet baryons have been studied from the 3-quark state picture in flavor SU(3) symmetry.
They belong to a 10-dimensional representation with similar masses
and the mass splittings can be also explained by the  Gell-Mann–Okubo mass formula.
In this picture,
the difference between $\Omega$ and $\Delta$ seems to be explained by the quark mass ($m_{q}$) dependence of the meson and baryon thresholds; the pseudoscalar meson masses are proportional to $\sqrt{m_{q}}$ while the baryon masses linearly depend on $m_{q}$.
On the other hand,
the difference between $\Omega$ and $\Delta$ become more non-trivial once we investigate these states
from the scattering theory with meson-baryon degrees of freedom (dof).
This approach, however,
could have a broader utility since it can handle a bound state and a resonance
on equal footing and applications to exotic hadrons are possible.
We thus reexamine the decuplet baryons from the aspects of meson-baryon dof
by first-principles lattice QCD calculations in this study. 
To this end, we extract the $\Xi\bar{K}$ potential for $\Omega$ and the  $N\pi$ potential for $\Delta$ in the HAL QCD method, and make a comparison between them. 
To reduce huge computational costs required for these scattering channels, 
we employ heavier quark masses, where $u, d$ quark masses are chosen at the value close to the $s$ quark mass with slightly broken SU(3) flavor symmetry. While both $\Omega$ and $\Delta$ appear as stable particles in this setup, an inequality $m_{N}+m_{\pi}-m_{\Delta}<m_{\Xi}+m_{\bar K}-m_{\Omega}$ still holds as seen in previous lattice QCD results~\cite{PACS-CS:2008bkb}, and we can investigate the rephrased question, ``what is the physical origin which brings this hierarchy?''.
As we will show later, the HAL QCD method is particularly useful for this purpose, since it can directly extract the interaction potentials and distinguish two possible origins of the hierarchy, one from the difference in interactions and another from difference in kinematics.

This paper is organized as follows. In Sec.~\ref{sec:intro_of_HAL}, we briefly review the HAL QCD method in meson-baryon systems. We define $N\pi$ and $\Xi\bar{K}$ 3-point correlation functions and their radial/spherical decompositions in Sec.~\ref{sec:MB3ptfunc}, and show simulation details in Sec.~\ref{sec:numsetup}. In Sec.~\ref{sec:results}, we present potentials for $N\pi$ and $\Xi\bar{K}$, explain the fitting procedure of the potentials, and show numerical results of observables such as scattering phase shifts, binding energies, and root-mean-square distances. Sec.~\ref{sec:conclusion} is devoted to a conclusion of this paper. Some technical details and additional contents related to our study are discussed in appendices.

\section{HAL QCD method in meson-baryon systems}\label{sec:intro_of_HAL}
We define a meson-baryon potential $U_{\alpha\beta}(\vb{r},\vb{r}')$ in the HAL QCD method~\cite{Ishii:2006ec, Aoki:2009ji} as
\begin{eqnarray} \label{eq:orig_idea_of_HAL}
 \int d^3r' \ U_{\alpha\beta}(\vb{r},\vb{r}') \Psi^{W}_{\beta}(\vb{r}') 
 = \Big( \frac{k^2}{2\mu} - H_{0} \Big) \Psi^{W}_{\alpha}(\vb{r}),
 \label{eq:defU}
\end{eqnarray}
where $W$ is a total energy, $k$ is a relative momentum in the center of mass frame, $\alpha, \beta$ are indices for upper spin components,
$\mu$ is a reduced mass and $H_{0}$ is a free Hamiltonian. The total energy $W$ is related to $k$ as $W=\sqrt{k^2+m_{M}^2}+\sqrt{k^2+m_{B}^2}$, where $m_{M}$ and $m_{B}$ are meson and baryon masses, respectively. 
An equal-time NBS wave function $\Psi^{W}_{\alpha}(\vb{r})$ is  defined by
\begin{eqnarray}\label{nbswf}
\Psi^{W}_{\alpha}(\vb{r}) =\bra{0} M(\vb{r} , 0) B_{\alpha}(\vb{0} , 0) \ket{MB;W},
\end{eqnarray}
where $\ket{0}$ is a vacuum state in QCD, $M(\vb{x},t)$ and $B_{\alpha}(\vb{x},t)$ are meson and baryon operators at spacetime $(\vb{x}, t)$, respectively, and $\ket{MB;W}$ is a meson-baryon state with energy $W$. The potential $U_{\alpha\beta}(\vb{r},\vb{r}')$ in Eq.(\ref{eq:orig_idea_of_HAL}) is energy-independent and non-local. 
Once  the potential $U_{\alpha\beta}(\vb{r},\vb{r}')$ is obtained,
we can extract an S-matrix for the meson-baryon scattering by solving the corresponding Schr\"{o}dinger equation.

In our study, we employ the time-dependent HAL QCD method~\cite{Ishii:2012ssm}, an improved version of the original HAL QCD method,
to extract potentials in lattice QCD. In this method, we first define an $R$-correlator as
\begin{eqnarray}\label{eq:defofRcorr}
R_{\alpha}(\vb{r},t) 
\equiv \frac{F_{\alpha}(\vb{r},t)}{C_{M}(t)C_{B}(t)},
\end{eqnarray}
where $C_{M}(t)$ and $C_{B}(t)$ are 2-point correlation functions for meson and baryon, respectively, and $F$ is a $n$-point correlation function ($n>2$) of the meson-baryon system, which is given by
\begin{eqnarray}
F_{\alpha}(\vb{r},t) = \bra{0}   M(\vb{r+x} , t+t_{0}) B_{\alpha}(\vb{x} , t+t_{0}) \ \bar{\mathcal{J}}_{MB} (t_{0}) \ket{0},
\end{eqnarray}
where $\bar{\mathcal{J}}_{M B} (t_{0})$ is the source operator at timeslice $t_{0}$, which creates meson-baryon states with the quantum numbers of interest from the vacuum.
The $R$-correlator can be decomposed into contributions from elastic and inelastic states as
\begin{eqnarray}\label{eq:propofRcorr}
R_{\alpha}(\vb{r},t)
 = \sum_{n}   A_{n} \Psi^{W_{n}}_{\alpha}(\vb{r}) \ e^{-\Delta  W_{n}t}+({\rm inelastic \ contributions}),
\end{eqnarray}
where $W_{n}$ is the energy of the $n$th eigen states, $A_{n}$ is a coefficient independent of $\vb{r}$  and $\alpha$, and $\Delta  W_{n} = W_{n}-m_{M}-m_{B}$ is an energy difference from the threshold. Eq.~\eqref{eq:defU} implies that 
each term in the elastic part, $A_{n} \Psi^{W_{n}}_{\alpha}(\vb{r})e^{-\Delta  W_{n}t}$, satisfies 
\begin{eqnarray}\label{eq:scheqofelastic}
 \left( \frac{k^2_{n}}{2\mu} - H_{0} \right) A_{n} \Psi^{W_{n}}_{\alpha}(\vb{r})e^{-\Delta  W_{n}t}
 =\int d^3r' \ U_{\alpha\beta}(\vb{r},\vb{r}') A_{n} \Psi^{W_{n}}_{\beta}(\vb{r'})e^{-\Delta  W_{n}t},
\end{eqnarray}
where $k_{n}^{2}/2\mu$ can be expressed in terms of $\Delta W_{n}$ as
\begin{eqnarray}\label{eq:ksqDeltaW}
\frac{k_{n}^{2}}{2\mu}
= \frac{ \mathcal{P}(\Delta W_{n})}{(\Delta W_{n}/M + 1)^2},
\end{eqnarray}
with $M=m_{M}+m_{B}$ and
\begin{eqnarray}\label{eq:polyofW}
\mathcal{P}(\Delta W_{n})
= \Delta W_{n} + \frac{\mu+M}{2\mu M}(\Delta W_{n})^2 + \frac{1}{2\mu M}(\Delta W_{n})^3 + \frac{1}{8\mu M^2}(\Delta W_{n})^4.
\end{eqnarray}
We thus express $k_n^2$ by an expansion in terms of $\Delta W_{n}$ as
\begin{eqnarray}
\begin{aligned}
\frac{k_{n}^2}{2\mu} 
= \Delta W_{n} + \frac{1+3\delta^2}{8\mu}(\Delta W_{n})^2 + \frac{M^2\delta^2}{8\mu}\sum_{k=3}^{\infty} (k+1)\left(\frac{-\Delta W_{n}}{M}\right)^k 
\equiv \sum_{k=1}^{\infty}C^{(k)}_{m_{M},m_{B}}(\Delta W_{n})^{k}, 
\end{aligned}
\end{eqnarray}
where $\delta = (m_{M}-m_{B})/M$. Rewriting $\Delta W_{n}$ in this series in terms of time derivatives and summing over $n$ in Eq.(\ref{eq:scheqofelastic}), we obtain
\begin{eqnarray}
\left[\sum_{k=1}^{\infty}C^{(k)}_{m_{M},m_{B}}\left(-\pdv{t}\right)^{k}-H_{0} \right]R_{\alpha}(\vb{r},t)
\simeq \int d^3r' \ U_{\alpha\beta}(\vb{r},\vb{r}')R_{\beta}(\vb{r'},t).
\label{eq:scheq_rela_approx}
\end{eqnarray}
for a large enough $t$ to suppress inelastic contributions.

Applying the Okubo-Marshak expansion~\cite{OKUBO1958166} to meson-baryon systems, the leading order (LO) term  of $U_{\alpha\beta}(\vb{r},\vb{r}')$ in the derivative expansion is given by
\begin{eqnarray}
U_{\alpha\beta}(\vb{r},\vb{r}') \simeq V^{\textrm{LO}}(r) \delta_{\alpha\beta}\delta^{(3)}(\vb{r}-\vb{r}'),
\end{eqnarray}
where $V^{\textrm{LO}}(r)$ can be extracted from $R_{\alpha}(\vb{r},t)$ for any $\alpha$ as
\begin{eqnarray}\label{eq:LOpotential_gen}
V^{\textrm{LO}}(r) \simeq
\frac{1}{R_{\alpha}(\vb{r},t)}\left[\sum_{k=1}^{\infty}C^{(k)}_{m_{M},m_{B}}\left(-\pdv{t}\right)^{k}-H_{0} \right]R_{\alpha}(\vb{r},t).
\end{eqnarray}
We truncate an infinite summation over $k$ by $k\le 2$ for $N\pi$ and $k\le 3$ for $\Xi \bar{K}$, respectively, since remaining higher order contributions are negligibly small\footnote{There is an alternative method to derive potentials without the expansion in $\Delta W$ by using at most the 3rd time derivatives, which is explained in Appendix~\ref{sec:exactrelhal}. We have confirmed that this exact one gives no significant differences from our results
for $N\pi$ and $\Xi \bar K$ potentials, showing that higher order contributions are indeed small.}.

\section{Correlation functions with single-baryon source operators}\label{sec:MB3ptfunc}

In order to investigate interactions of P-wave $I=3/2$ $N\pi$ and $I=0$ $\Xi\bar{K}$ systems in the HAL QCD method at low energies,
we use following 3-point correlation functions,  
\begin{eqnarray}
F^{N\pi}_{\alpha,j_{z}}({\bf r},t) 
&=& \bra{0} \pi^{+}({\bf r+x},t) p_{\alpha}({\bf x},t) \ \bar{\Delta}^{++}_{j_{z}}(t_{0})\ket{0}, \label{eq:def_of_Npi3pt}\\
F^{\Xi\bar{K}}_{\alpha,j_{z}}({\bf r},t) 
&=& \bra{0} \frac{1}{\sqrt{2}}(K^{-}({\bf r+x},t) \Xi^{0}_{\alpha}({\bf x},t)-\bar{K}^{0}({\bf r+x},t) \Xi^{-}_{\alpha}({\bf x},t)) \  \bar{\Omega}^{-}_{j_{z}}(t_{0})\ket{0}, \label{eq:def_of_XiKbar3pt}
\end{eqnarray}
where sink operators are defined by
\begin{eqnarray}
\pi^{+}(x) &=& -i\bar{d}(x)\gamma_{5}u(x), \quad 
\bar{K}^{0}(x) = -i\bar{d}(x)\gamma_{5}s(x), \quad 
K^{-}(x) = i\bar{u}(x)\gamma_{5}s(x), \\
p_{\alpha}(x) &=&\epsilon_{abc}u_{a,\alpha}(x)(u^{{\textrm T}}_{b}(x) C\gamma_{5}d_{c}(x)), \\ 
\Xi^{0}_{\alpha}(x) &=&\epsilon_{abc}s_{a,\alpha}(x)(s^{{\textrm T}}_{b}(x) C\gamma_{5}u_{c}(x)) , \quad
\Xi^{-}_{\alpha}(x) =\epsilon_{abc}s_{a,\alpha}(x)(s^{{\textrm T}}_{b}(x) C\gamma_{5}d_{c}(x)).
\end{eqnarray}
Since we expect bound decuplet baryons to appear below thresholds, we employ 3-quark-type decuplet baryon operators at the source, where the explicit forms are given by
\begin{eqnarray}\label{eq:def_of_dec_bary_op}
\begin{aligned}
&D_{+\frac{3}{2}}(t_{0}) 
= \sum_{{\bf z}}\epsilon_{abc} (q^{T}_{b}({\bf z},t_{0}) \Gamma_{+} q_{c}({\bf z},t_{0}))q_{a,0}({\bf z},t_{0}), \\
&D_{+\frac{1}{2}}(t_{0}) 
= \frac{1}{\sqrt{3}}\sum_{{\bf z}}\epsilon_{abc} 
[\sqrt{2}(q^{T}_{b}({\bf z},t_{0}) \Gamma_{z} q_{c}({\bf z},t_{0}))q_{a,0}({\bf z},t_{0}) 
+(q^{T}_{b}({\bf z},t_{0}) \Gamma_{+} q_{c}({\bf z},t_{0}))q_{a,1}({\bf z},t_{0})
], \\
&D_{-\frac{1}{2}}(t_{0}) 
= \frac{1}{\sqrt{3}}\sum_{{\bf z}}\epsilon_{abc}
[\sqrt{2}(q^{T}_{b}({\bf z},t_{0}) \Gamma_{z} q_{c}({\bf z},t_{0}))q_{a,1}({\bf z},t_{0}) 
+(q^{T}_{b}({\bf z},t_{0}) \Gamma_{-} q_{c}({\bf z},t_{0}))q_{a,0}({\bf z},t_{0})
], \\
&D_{-\frac{3}{2}}(t_{0}) 
= \sum_{{\bf z}}\epsilon_{abc} 
(q^{T}_{b}({\bf z},t_{0}) \Gamma_{-} q_{c}({\bf z},t_{0}))q_{a,1}({\bf z},t_{0}),
\end{aligned}
\end{eqnarray}
with $\Gamma_{\pm} = \frac{1}{2}C(\gamma_{2} \pm i\gamma_{1})$ and $\Gamma_{z} = \frac{-i}{\sqrt{2}}C\gamma_{3}$, and $q=(u, s)$ for $D=(\Delta^{++}, \Omega^-)$, respectively. Indeed each operator in Eq.~(\ref{eq:def_of_dec_bary_op}) couples to a spin-3/2 particle with a different $z$ component of the spin, since they belong to an $H_{g}$ irreducible representation of the cubic group $O^{D}_{h}$. A summation over ${\bf z}$ selects zero total momentum states.

To obtain NBS wave functions with $J^{P}=3/2^{+}$, we project $F_{\alpha,j_{z}}({\bf r},t)$ onto the same component in the $H_{g}$ representation of $\bar{D}_{j_{z}}(t_{0})$. Using Clebsch--Gordan coefficients, projected 3-point correlation functions can be described as
\begin{eqnarray}\label{eq:partdec3pt}
\begin{aligned}
\mqty(
F_{\uparrow,+\frac{3}{2}}(\textbf{r}, t) \\
F_{\downarrow,+\frac{3}{2}}(\textbf{r}, t)
)
&=
f_{+\frac{3}{2}}(r,t)
\mqty(
Y_{1,+1}(\vb{\hat{r}}) \\
0  
), \quad
\mqty(
F_{\uparrow,+\frac{1}{2}}(\textbf{r}, t) \\
F_{\downarrow,+\frac{1}{2}}(\textbf{r}, t)
) 
=
f_{+\frac{1}{2}}(r,t)
\mqty(
\sqrt{\frac{2}{3}}Y_{1,0}(\vb{\hat{r}}) \\
\sqrt{\frac{1}{3}}Y_{1,+1}(\vb{\hat{r}})  
), \\
\mqty(
F_{\uparrow,-\frac{1}{2}}(\textbf{r}, t) \\
F_{\downarrow,-\frac{1}{2}}(\textbf{r}, t)
)  
&=
f_{-\frac{1}{2}}(r,t)
\mqty(
\sqrt{\frac{1}{3}}Y_{1,-1}(\vb{\hat{r}}) \\
\sqrt{\frac{2}{3}}Y_{1,0}(\vb{\hat{r}})  
), \quad
\mqty(
F_{\uparrow,-\frac{3}{2}}(\textbf{r}, t) \\
F_{\downarrow,-\frac{3}{2}}(\textbf{r}, t)
)   
=
f_{-\frac{3}{2}}(r,t)
\mqty(
0 \\
Y_{1,-1}(\vb{\hat{r}})  
),
\end{aligned}
\end{eqnarray}
where $Y_{l,m}(\vb{\hat{r}})$ is the spherical harmonics and $f_{j_{z}}(r,t)$ is a factor that depends only on $r=|\vb{r}|$ and $t$. 
We extract $f_{j_{z}}(r,t)$ using a projection to $(l=1,m)$, defined on a discrete space as
\begin{eqnarray}\label{eq:descproj}
f_{j_{z}}(r,t) = \frac{\sum_{\vb{r'}\in \{\vb{r'}|r'=r \}}Y^{*}_{1m}(\vb{\hat{r}}')F_{\alpha,j_{z}}(\vb{r'}, t)}{\sum_{\vb{r'}\in \{\vb{r'}|r'=r \}}Y^{*}_{1,m}(\vb{\hat{r}}')Y_{1m}(\vb{\hat{r}}')}
\end{eqnarray}
with corresponding $(m, \alpha)$ for each $j_{z}$. For $j_{z} = \pm 1/2$, we can derive $f_{j_{z}}(r,t)$ in two ways by setting either $(m, \alpha) = (0,\uparrow)$ or $(1,\downarrow)$ for $j_{z} = +1/2$ and either $(m, \alpha) = (0,\downarrow)$ or $(-1,\uparrow)$ for $j_{z} = -1/2$, respectively. In this paper, we take an average over the factors calculated from the two choices.

Applying Eq.(\ref{eq:partdec3pt}) to Eq.(\ref{eq:defofRcorr}) and Eq(\ref{eq:LOpotential_gen}), we obtain
\begin{eqnarray}\label{eq:LOpotential_dec}
V^{\textrm{LO}}(r) \simeq
\frac{1}{\mathcal{R}_{j_{z}}(r,t)}\left[\sum_{k=1}^{\infty}C^{(k)}_{m_{M},m_{B}}\left(-\pdv{t}\right)^{k}
+\frac{1}{2\mu}\left(\frac{1}{r}\pdv[2]{r}r  - \frac{l(l+1)}{r^2}\right) \right]\mathcal{R}_{j_{z}}(r,t),
\end{eqnarray}
with the angular momentum $l=1$ and
\begin{eqnarray}
\mathcal{R}_{j_{z}}(r,t)= \frac{f_{j_{z}}(r,t)}{C_{M}(t)C_{B}(t)}.
\end{eqnarray}
In this study, we use this equation to extract the LO potentials. 
Since $f_{j_{z}}(r,t)$ for any $j_{z}$ gives the same LO potential thanks to a rotation symmetry, we take an average over $j_{z}$
to increase statistics.
Furthermore, a charge conjugation symmetry provides a relations among $f_{j_{z}}(r,t)$ as
\begin{eqnarray}\label{ccofraddirof3pt}
\begin{aligned}
 f_{+\frac{3}{2}}(r,t) = -f^{\ast}_{-\frac{3}{2}}(r,t), \quad
f_{+\frac{1}{2}}(r,t) = -f^{\ast}_{-\frac{1}{2}}(r,t),
\end{aligned}
\end{eqnarray}
so that an average of $f_{j_{z}}(r,t)$ over $j_{z}$ is guaranteed to be pure imaginary, and we therefore ignore its real part.

\section{Simulation details}\label{sec:numsetup}

In our numerical calculations, we employ (2+1)-flavor gauge configurations generated by the PACS-CS Collaboration with the improved Iwasaki gauge action and the $\order{a}$-improved Wilson quark action at $\beta = 1.90$ on $32^3 \times 64$ lattice~\cite{PACS-CS:2008bkb}, which corresponds to $0.0907(13)$~fm for the lattice spacing. Hopping parameters of the ensemble in our calculations are $\kappa_u = \kappa_d=0.13754$ and $\kappa_{s}=0.13640$. A periodic boundary condition is imposed in all spacetime directions. We use 450 configurations with 16 sources at different time slices on each configuration, and average forward and backward propagations to increase statistics. Statistical errors are estimated by the jackknife method with a binsize of 45 configurations.

We employ a smeared quark source with an exponential smearing function~\cite{Iritani:2016jie}, which is given by
\begin{eqnarray}
 S_{A,B}(\vb{x}) =
 \begin{cases}
A e^{-B|\vb{x}|} & (|\vb{x}|<\frac{L-1}{2}) \\
1 & (|\vb{x}|=0) \\
0 & (|\vb{x}| \geq\frac{L-1}{2})
 \end{cases} \label{eq:smearingfunction}
\end{eqnarray}
in lattice unit, where we take $(A,B)=(1.2, 0.17)$ for light quarks and $(A,B)=(1.2, 0.25)$ for the strange quark. We also apply the same smearing to quarks at the sink with $(A,B)=(1.0, 1/0.7)$ to reduce singular behaviors of potentials at short distances~\cite{Akahoshi:2021sxc}. Details of singular behaviors and associated issues for its fitting are explained in Appendix~\ref{sec:ss}.

\begin{figure}
    \centering
    \includegraphics[width=1.0\textwidth]{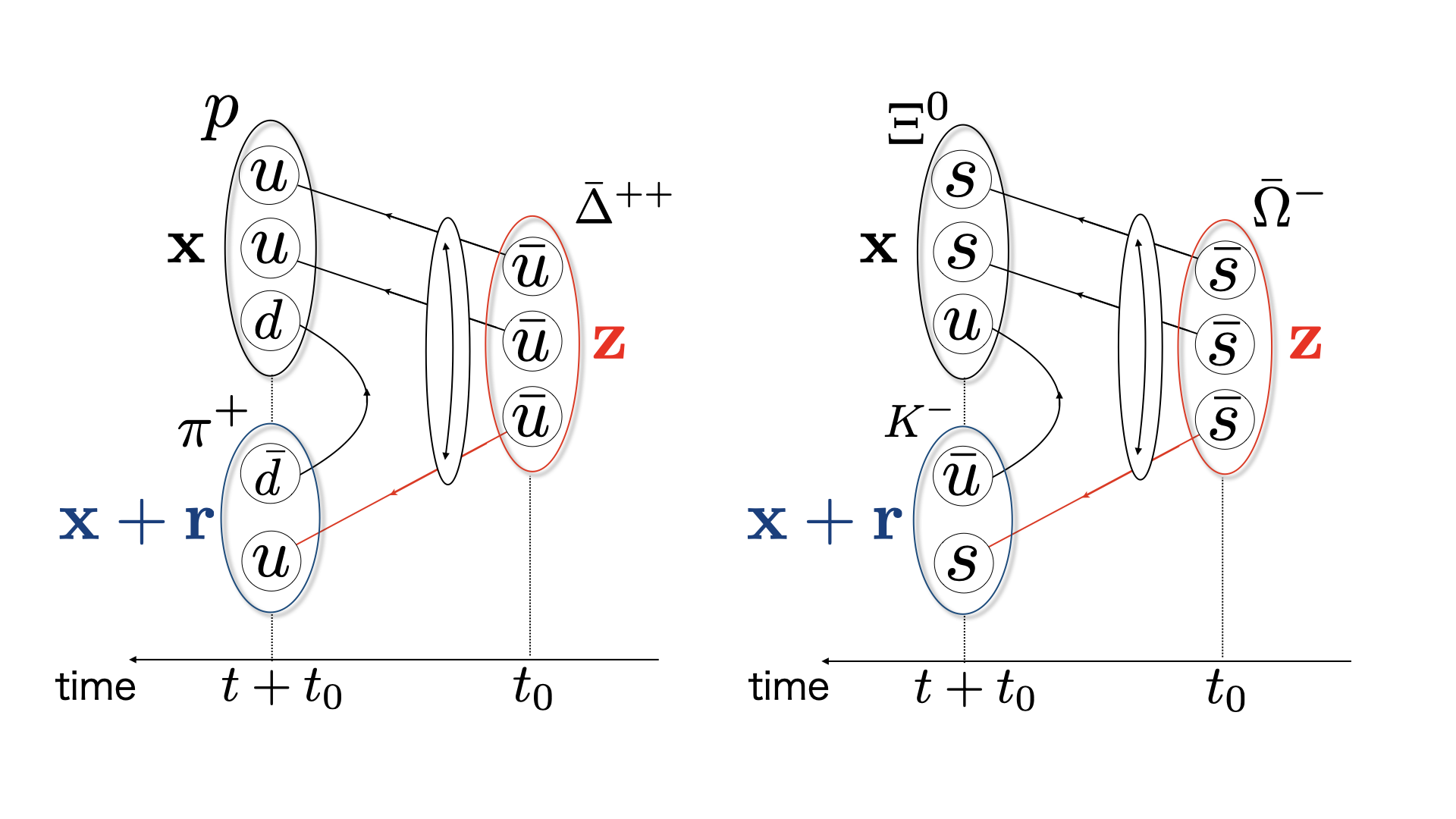}
    \caption{Quark contraction diagram corresponding to Eq.(\ref{eq:def_of_Npi3pt}) (Left) and the first term in Eq.(\ref{eq:def_of_XiKbar3pt}) (Right). A circle with two-way arrow across three lines indicates permutations of quark contractions among them. All-to-all propagators are used in red lines, while point-to-all propagators are used in black lines.}
    \label{fig:contraction}
\end{figure}

We show quark contraction diagrams corresponding to Eq.(\ref{eq:def_of_Npi3pt}) and the first term of Eq.(\ref{eq:def_of_XiKbar3pt}) in Fig.\ref{fig:contraction} (Left)  and  Fig.\ref{fig:contraction} (Right), respectively. 
The second term of Eq.(\ref{eq:def_of_XiKbar3pt})  is obtained from the first one  by replacing $u$ with $d$,
and thus gives a contribution identical to the first one with $m_{u}=m_{d}$.
The spatial coordinate ${\bf z}$ at the source is summed over so that source operators have zero momentum. In this case, quark propagators represented by red lines in this figure are all-to-all propagators. We use the conventional stochastic technique to estimate them approximately, together with dilutions~\cite{Foley:2005ac} for color/spinor/time components and the s2 (even/odd) dilution~\cite{Akahoshi:2019klc} for the position ${\bf z}$.

For the spatial coordinate ${\bf x}$ at the sink, the calculation is performed with fixed ${\bf x}$, so that quark propagators represented by black lines are obtained by point-to-all propagators which requires smaller numerical cost. In order to increase the statistics using the translational invariance, we calculate at 64 different spatial points given by ${\bf x}+\Delta{\bf x}$ with $\Delta{\bf x} = (0,0,0), (0,0,8), \cdots, (24,24,24)$. As a method which keeps this computational cost moderate, we employ the truncated solver method~\cite{Bali:2009hu} combined with the covariant-approximation averaging~\cite{Shintani:2014vja}, namely an all-mode averaging (AMA) without low modes. For the specific value of ${\bf x}$, we choose it randomly for each gauge configuration to keep the exact covariance of the AMA, which might be broken by round-off errors. See Appendix C in Ref.~\cite{Shintani:2014vja}.

Meson 2-point correlation functions are calculated using all-to-all propagators, where the one-end trick~\cite{Foster:1998vw, McNeile:2002fh} is employed. Baryon 2-point correlation functions are calculated using point-to-all propagators. Also, both in the meson and baryon 2-point functions, we employ the smearing to quarks at the source and the sink with the function given in Eq.~(\ref{eq:smearingfunction}). Masses of single hadrons obtained from them in this setup are listed in Table \ref{tab:hadronmass}. Since $\Delta$ and $\Omega$ masses lie below $N\pi$ and $\Xi\bar{K}$ threshold energies, respectively, they appear as bound states in this setup.

\begin{table}

\begin{ruledtabular}
\begin{tabular}{c|cccccc}
hadron&
$\pi$ &
$K$ &
$N$ &
$\Xi$ &
$\Delta$ &
$\Omega$ 
 \\
\colrule
mass &  411.2(1.7) & 635.1(1.5)& 
           1217.2(4.7)& 1505.3(4.5)&
            1522.9(7.8)& 1847.0(6.5) 
\\
fit range &  [10,30] & [10,30]& 
           [7,20]& [7,20]&
            [6,15]& [6,20]
\end{tabular}
\end{ruledtabular}
\caption{\label{tab:hadronmass}%
Hadron masses in MeV estimated by fitting  2-point functions. The second row shows temporal fitting ranges in lattice unit.}
\end{table}

\section{P-wave $N\pi$ and $\Xi\bar{K}$ interactions}\label{sec:results}
 
\subsection{Potentials}
\begin{figure}[t]
    \begin{center}
        \includegraphics[width=0.49\textwidth]{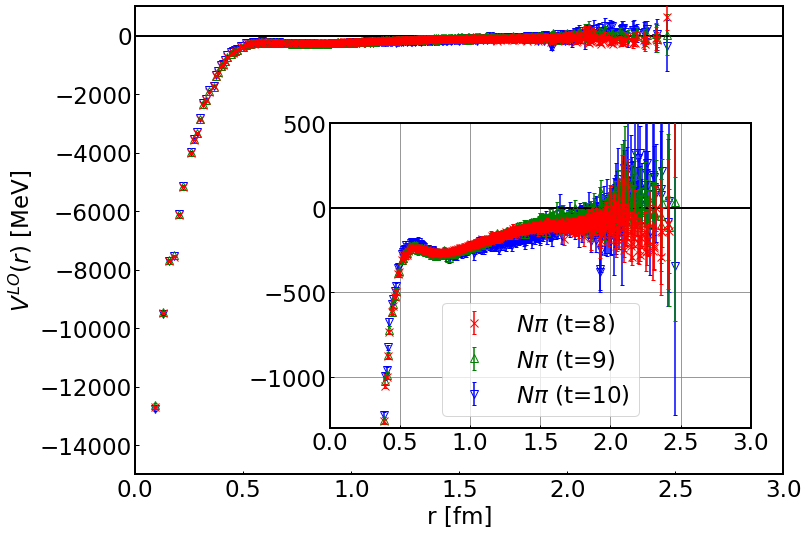}
	    \includegraphics[width=0.48\textwidth]{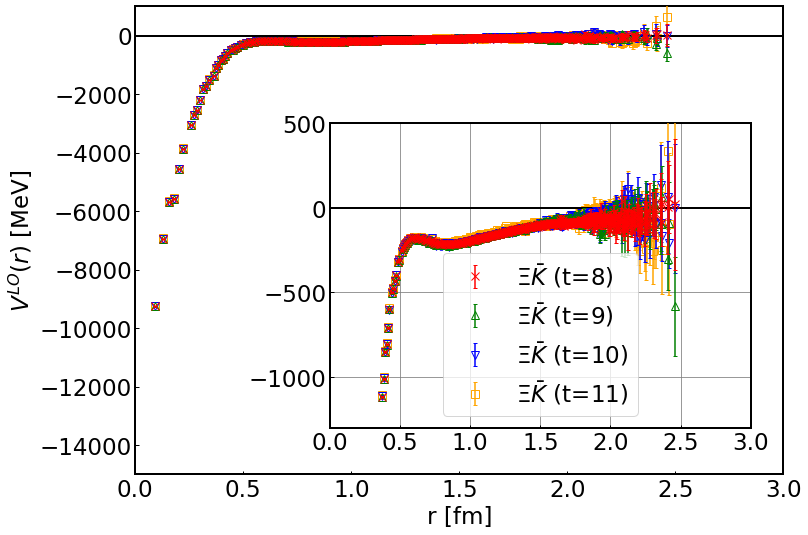}
	    \caption{The leading-order potentials of the $N\pi$ system at $t=8$--$10$ (Left) and the $\Xi\bar{K}$ system at $t=8$--$11$ (Right). The laplacian terms are calculated in 4th order accuracies.}
	    \label{fig:rawpotdata}    
    \end{center}
\end{figure}

In Fig.~\ref{fig:rawpotdata}, we present LO potentials for P-wave $N\pi$ at $t=8$--$10$ and $\Xi\bar{K}$ at $t=8$--$11$, 
where effective masses of $\Delta$ and $\Omega$ reach plateaux, respectively.
We do not observe significant $t$-dependence of potentials, indicating that inelastic contributions as well as effects of higher-order terms in the derivative expansion of the potential are well under control.

Both potentials have very strong attractions at short distances, which can be responsible to produce bound states corresponding to $\Delta$ and $\Omega$. 
We also observe that the attraction of the $N\pi$ potential is deeper than that of the $\Xi\bar{K}$ by a few thousands MeV at short distances.

Both potentials have similar shapes at middle and long distances, where meson-baryon interactions may be dominated by one meson exchanges. One possibility is that the $N\pi$ system exchanges a $\rho$ meson while the $\Xi\bar{K}$ system exchanges a $\rho$ meson or an octet part of $\phi$/$\omega$, where the masses of these vector mesons almost degenerate in our lattice setup near the SU(3) flavor symmetric point ($m_{\rho}/m_{\phi}\approx 0.80$)~\cite{PACS-CS:2008bkb}\footnote{Note that this discussion is only qualitative because even the Compton wave length of the lightest pion is $0.5$~fm in this setup.}. If this is a relevant picture, two potentials at middle and long distances remain to be similar even at the physical point, since mass difference of vector mesons are not so large at the physical point ($m_{\rho}/m_{\phi}\approx 0.75$).

Although both potentials go to zero within errors at long distances, interaction ranges are longer than 
a half of the box size, $L/2 \approx 1.45$ fm. We therefore carefully check possible finite-volume effects on observables
  as will be explained below.

\subsection{Systematic uncertainties of the potentials and fitting results}

We estimate systematic uncertainties in our analysis as follows.

Finite-volume effects are estimated by using two types of fit functions,  one is a simple three Gaussians as
\begin{eqnarray}\label{eq:3G}
V^{3G}(\vb{r}) 
= a_{0}e^{-(r/a_{1})^2}+a_{2}e^{-(r/a_{3})^2}+a_{4}e^{-(r/a_{5})^2}, 
\end{eqnarray}
where we assume that $a_{1} < a_{3} < a_{5}$, and the other is a three Gaussians with its 6 mirror images~\cite{Akahoshi:2020ojo} as
\begin{eqnarray}\label{eq:3G_PBC}
V^{3G}_{P}(\vb{r})
= V^{3G}(\vb{r}) + \sum_{\vb{n}}V^{3G}(\vb{r}+L\vb{n}),
\end{eqnarray}
where $\vb{n} \in \{ (0,0,\pm1),(0,\pm1,0),(\pm1,0,0) \}$. 
We then employ only $V^{3G}(\vb{r})$ in both cases for the Schr\"{o}dinger equations to be solved.

We estimate effects of the leading order truncation for non-localities,
by employing potentials at different $t$, from $t=8$ to $10$ for $N\pi$ and from $t=8$ to $11$ for $\Xi\bar{K}$,
since $t$ dependences of potentials are mainly caused by the truncation of the derivative expansion.

Finite lattice spacing effects are estimated by two approaches. In the first approach, we evaluate the difference between the laplacian term in Eq.~(\ref{eq:LOpotential_gen}) calculated with 2nd and 4th order accuracies, whose explicit forms are given by
\begin{eqnarray}
\begin{aligned}
(\nabla^{2}R(\vb{r}))_{\textrm{2nd}} 
&= \sum_{{\bf i}={\bf \hat{x}},{\bf \hat{y}},{\bf \hat{z}}}\frac{R(\vb{r}+a{\bf i})-2R(\vb{r})+R(\vb{r}-a{\bf i})}{a^2}, \\
(\nabla^{2}R(\vb{r}))_{\textrm{4th}} 
&= \sum_{{\bf i}={\bf \hat{x}},{\bf \hat{y}},{\bf \hat{z}}}\frac{-R(\vb{r}+2a{\bf i})+16R(\vb{r}+a{\bf i})-30R(\vb{r})+16R(\vb{r}-a{\bf i})-R(\vb{r}-2a{\bf i})}{12a^2}.
\end{aligned}
\end{eqnarray}
In the second approach, we estimate the uncertainty from the difference between fits with and without data at $r=a$. For the fit including data at $r=a$, we find that modification for fit functions is necessary. In fact, if we naively fit with Eqs.~(\ref{eq:3G}), (\ref{eq:3G_PBC}), the fit results fail to reproduce original data point at $r=a$, probably because potentials with a strong attraction at short distances and a non-monotonic behavior at middle distances are too complicated to be fitted by three Gaussians. Similar difficulties appear also for other functions such as four Gaussians or two Gaussians plus one Yukawa potential.
To avoid this difficulty, we employ an interpolation 
combined with the usual fitting in the following way.
First, we fit potentials
excluding data at $r=a, \sqrt{2}a$. Then we perform a quadratic interpolation using original data at $r=a, \sqrt{2}a$ and the fit result at $r=(\sqrt{3}a+2a)/2$. For calculations of observables, we use a combination of fit results in $\left[(\sqrt{3}a+2a)/2,\infty \right)$ and the interpolation in $\left(0, (\sqrt{3}a+2a)/2\right]$. While only a continuity is guaranteed but a smoothness is not at $r=(\sqrt{3}a+2a)/2$ in this method, we do not find any serious non-smoothness in the results.

In summary, for potentials at a given $t$, there are $2 \times 2\times 2 = 8$ combinations of fitting schemes, which are listed in Table.\ref{tab:fit_list}. 

\begin{table}
\begin{tabularx}{0.7\textwidth}{ 
   >{\centering\arraybackslash}X 
  | >{\centering\arraybackslash}X 
   >{\centering\arraybackslash}X 
   >{\centering\arraybackslash}X}
  
& fit potential
& data at $r=a$ 
& accuracy of $\nabla^2$
 \\
\hline\hline
Fit 1 &  $V^{3G}(\vb{r})$& not included & 2nd order \\
Fit 2 &  $V^{3G}(\vb{r})$& not included & 4th order \\
Fit 3 &  $V^{3G}(\vb{r})$& included & 2nd order \\
Fit 4 &  $V^{3G}(\vb{r})$& included & 4th order \\
Fit 5 &  $V^{3G}_{P}(\vb{r})$& not included & 2nd order \\
Fit 6 &  $V^{3G}_{P}(\vb{r})$& not included & 4th order \\
Fit 7 &  $V^{3G}_{P}(\vb{r})$& included & 2nd order \\
Fit 8 &  $V^{3G}_{P}(\vb{r})$& included & 4th order \\
\end{tabularx}
\caption{\label{tab:fit_list}
A list of combinations of fit functions and potential data to estimate systematic uncertainties.
}
\end{table}

Central values of physical observables are calculated from Fit 2 at $t=9$ for $N\pi$ and at $t=10$ for $\Xi\bar{K}$. Fit parameters are listed in Table~\ref{tab:fitparam_central}, where $\chi^2/dof=4.5$ for $N\pi$ and $\chi^2/dof=36.0$ for $\Xi\bar{K}$, where we employ uncorrelated fit. This large $\chi^2/dof$ comes mostly from deviations between fit results and original data at short distances, which have small statistical errors and large systematic uncertainties,
even though most problematic data at $r=a$ are excluded for this Fit 2. Systematic errors are estimated from results in other fitting schemes and at other timeslices. 

\begin{table}
\begin{ruledtabular}
\begin{tabular}{c|ccccccccc}
system    &
$a_{0}$ [MeV]&
$a_{1}$ [fm]&
$a_{2}$ [MeV]&
$a_{3}$ [fm]&
$a_{4}$ [MeV]&
$a_{5}$ [fm]&

 \\
\colrule
$N\pi$   &  -13311.7(46.2) & 0.24(0.00)
            & 693.8(197.9) & 0.56(0.08)
            & -615.3(217.7) & 1.08(0.14), \\
$\Xi\bar{K}$  &  -9651.8(12.1) & 0.24(0.00)& 
            462.0(67.7) & 0.60(0.04) &
            -427.9(72.4) & 1.25(0.10) \\
\end{tabular}
\end{ruledtabular}
\caption{\label{tab:fitparam_central}%
Fit parameters $\{ a_{n} \}$ for potential data in Fit 2 at $t=9$ for $N\pi$ and at $t=10$ for $\Xi\bar{K}$, used to calculate
central values of observables.
}
\end{table}

\begin{figure}[htbp]
    \begin{center}
        \includegraphics[width=0.49\textwidth]{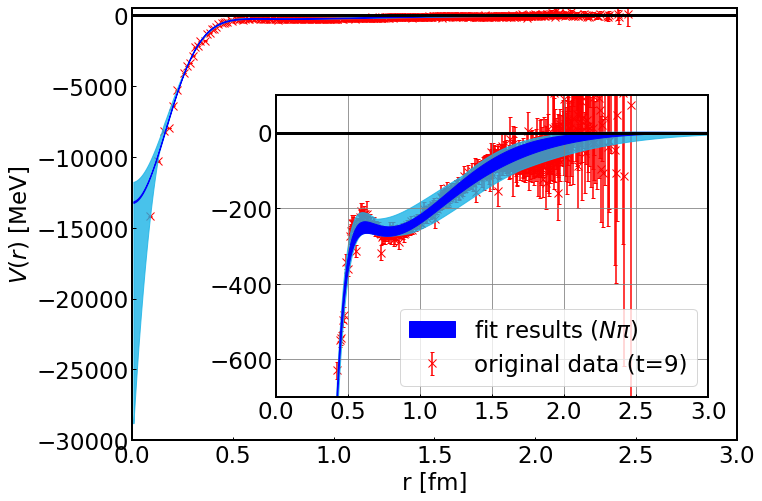}
	    \includegraphics[width=0.48\textwidth]{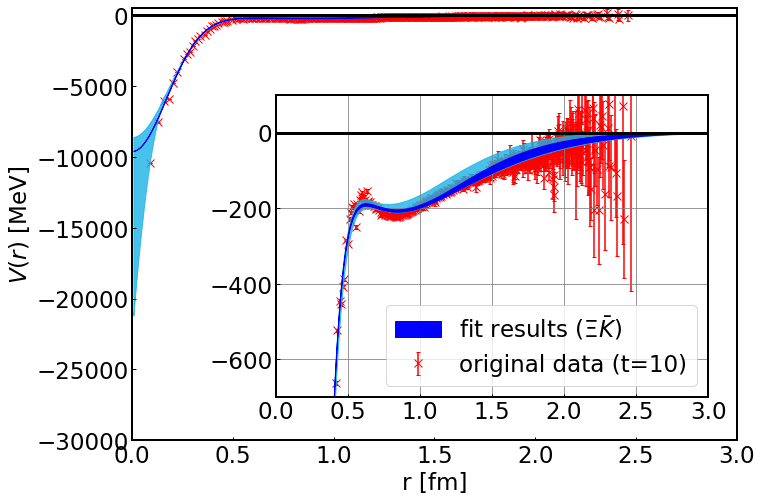}
	    \caption{Fit results for $N\pi$ (Left) and $\Xi\bar{K}$ (Right). Dark (light) blue bands show statistical errors (statistical and systematic errors added in quadrature), where systematic errors are estimated from other fitting schemes and $t$ dependences.
        Red crosses represent original potential data.}
	    \label{fig:Npi_potfit}    
    \end{center}
\end{figure}

Fig.~\ref{fig:Npi_potfit} presents original data by red crosses and fitting results by blue bands. The dark blue bands correspond to statistical errors in Fit 2 at $t=9$ for $N\pi$ and at $t=10$ for $\Xi\bar{K}$, whereas light blue bands show statisitcal and systematic errors added in quadrature, where the latter error is estimated from other fitting schemes and $t$ dependences. Note that systematic uncertainties at short distances, mainly caused by finite lattice spacing effects, are much larger than those at long distances caused by finite-volume effects. 

\begin{figure}
    \begin{center}
        \includegraphics[width=0.49\textwidth]{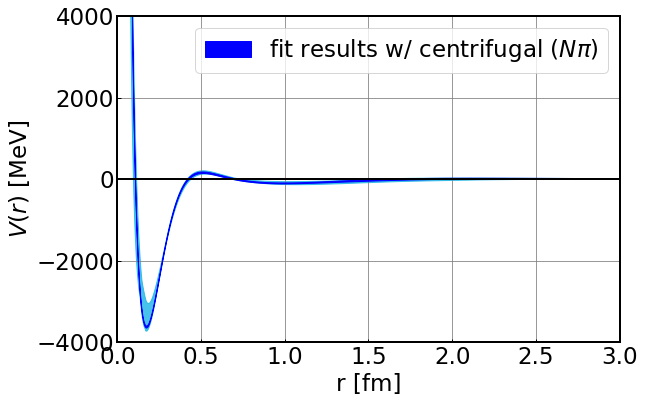}
	    \includegraphics[width=0.48\textwidth]{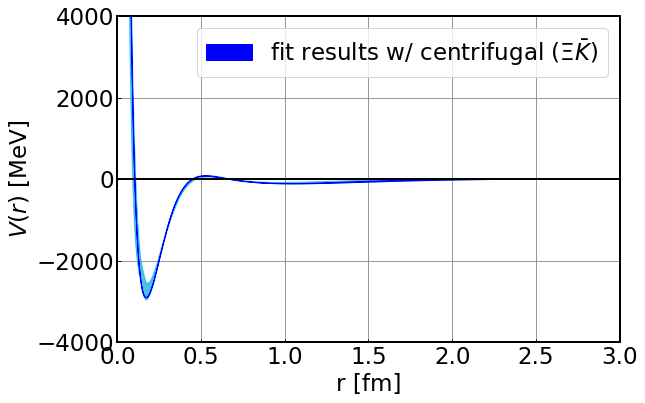}
	    \caption{A fit result with a centrifugal term for $N\pi$ (Left) and $\Xi\bar{K}$ (Right).
        Dark and light blue bands represent statistical errors
          and statistical and systematic errors added in quadrature, respectively.
        }
	    \label{fig:Npi_potfit_cent}    
    \end{center}
\end{figure}

Fig.~\ref{fig:Npi_potfit_cent} shows the fitted potential with a centrifugal term for $N\pi$ (Left) and $\Xi \bar K$ (Right). 
Both have an attractive pocket with a depth of about $3~\textrm{GeV}$
at $r\approx 0.2~\textrm{fm}$, a barrier with a height of about a hundred MeV at $r\approx 0.5~\textrm{fm}$, 
and a shallow pocket with a depth of about a hundred MeV at $r\approx 1.0~\textrm{fm}$. 

\subsection{Phase shifts, binding energies, and root-mean-square distances}
We solve the Schr\"{o}dinger equation in the radial direction as
\begin{eqnarray}\label{eq:scheq_radial}
 -\frac{1}{2\mu}\Big(\frac{1}{r}\dv[2]{r}r - \frac{l(l+1)}{r^2}\Big)\psi^{l,E}_{R}(r) + V(r)\psi^{l,E}_{R}(r) = \frac{k^2}{2\mu}\psi^{l,E}_{R}(r),
\end{eqnarray}
where $k$ is the relative momentum, $V(r)$ is the fitted potential, and the angular momentum is fixed to $l=1$ for the P-wave scattering. The total energy $E$ is related to $k^2$ as $E=\sqrt{k^2+m_{M}^2}+\sqrt{k^2+m_{B}^2}$. We then extract a scattering phase shift from its solution. 
Finally, by varying $k^2$ (or $E$), we determine an energy dependence of the scattering phase shift.

\begin{figure}[t]
    \begin{center}
        \includegraphics[width=0.49\textwidth]{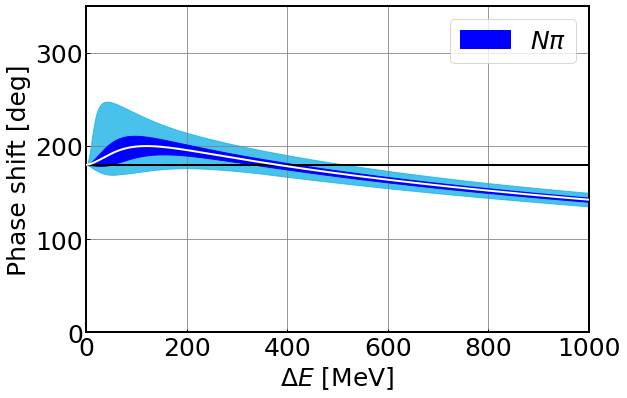}
	    \includegraphics[width=0.48\textwidth]{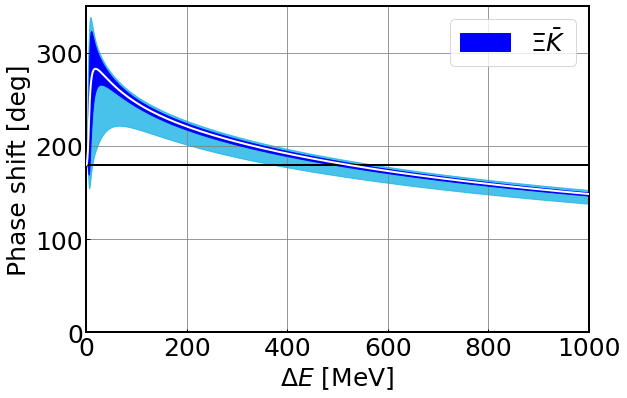}
	    \caption{Scattering phase shifts for $N\pi$ (Left) and $\Xi\bar{K}$ (Right).
        Dark and light blue bands represent statistical errors
          and statistical and systematic errors added in quadrature, respectively.
        }
	    \label{fig:phase}    
    \end{center}
\end{figure}

Fig.~\ref{fig:phase} presents the scattering phase shift in both channels as a function of energy measured from the 2-body threshold, $\Delta E = E -m_{M} - m_{B}$.
Both results show attractive behaviors at low energies, as suggested by shapes of potentials.
As we increase energy further,  the scattering phase shift in each channel approach zero, suggesting an existence of one bound state according to
the Levinson's theorem.

\begin{figure}[t]
    \centering
    \includegraphics[width=1.0\textwidth]{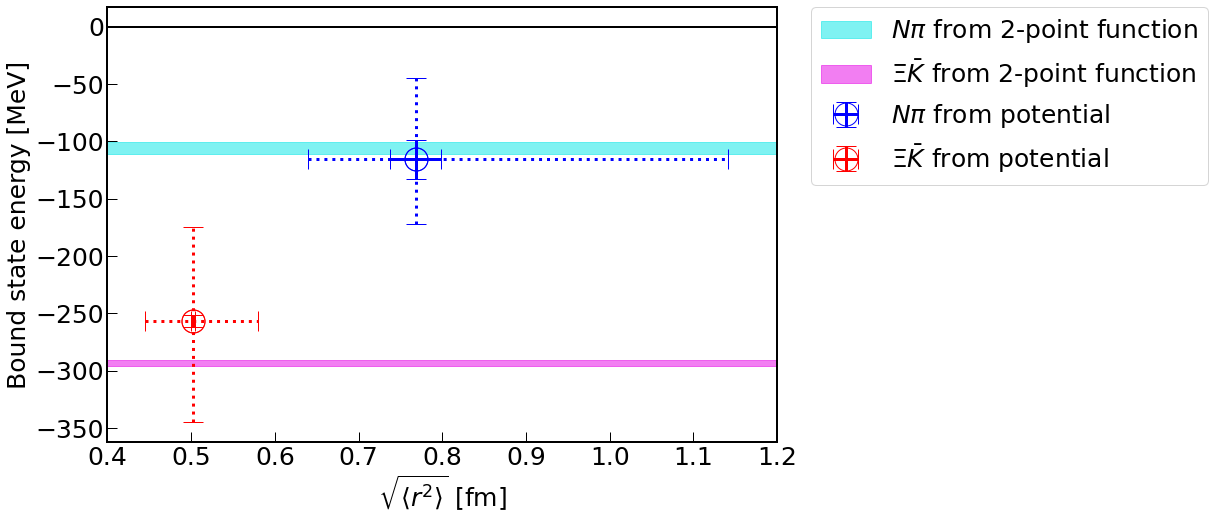}
    \caption{A binding energy (vertical axis) and a root-mean-square distance (horizontal axis) for $N\pi$ (blue) and $\Xi\bar{K}$ (red). Solid and dotted error bars show statistical errors and statistical and systematic errors added in quadrature, respectively. Cyan and magenta bands show binding energies estimated from $\Delta$ and $\Omega$ 2-point functions, respectively. Note that there is no constraint for root-mean-square distances from 2-point functions.
    }
    \label{fig:be}
\end{figure}

We calculate a binding energy in each channel by solving the Schr\"{o}dinger equation via the Gaussian Expansion Method (GEM)~\cite{HIYAMA2003223}, which gives
\begin{eqnarray}
 E^{N\pi}_{\textrm{bind}} &=& 115.6 (17.2) \smqty(+54.0 \\ -69.3)~\mbox{MeV}, \\
 E^{\Xi\bar{K}}_{\textrm{bind}} &=& 256.6 (5.5) \smqty(+88.2 \\ -82.2)~\mbox{MeV},
\end{eqnarray}
where first and second errors represent statistical and systematic errors, respectively. 
As shown in Fig.~\ref{fig:be}, binding energies in both channels are consistent with the estimate using baryon masses measured from 2-point functions
as $m_{N}+m_{\pi}-m_{\Delta} = 105.5(5.2)~\textrm{MeV}$ and  $m_{\Xi}+m_{\bar{K}}-m_{\Omega} = 293.5(2.8)~\textrm{MeV}$
within large systematic errors.

The binding energy in the $\Xi \bar K$ system is larger than that in the $N\pi$ system by more than a hundred MeV. Combined with the previous observation that the $\Xi \bar K$ potential has weaker attraction than the $N\pi$, it is indicated that a reason for a larger binding energy in the $\Xi \bar K$ than in the $N\pi$ is not the difference of the two interactions.

We also calculate root-mean-square distances of wave functions obtained via the GEM for bound states, which represent the distances between meson and baryon components in $\Delta$ and $\Omega$ baryons
\footnote{
    Note that the root-mean square distance defined here is different from usual charge/matter root-mean square distance.
}.
We obtain 
\begin{eqnarray}
 \sqrt{\langle r^2 \rangle_{N\pi}}  &=& 0.77 (0.03)
\smqty(+0.37 \\ -0.12)~\mbox{fm},\\
 \sqrt{\langle r^2 \rangle_{\Xi\bar{K}}} &=& 0.50 (0.00)
\smqty(+0.08 \\ -0.06)~\mbox{fm},
\end{eqnarray}
where first and second errors represent statistical and systematic errors, respectively. 
Results are also shown in Fig.~\ref{fig:be} (horizontal axis). 
Sizes of both bound states estimated by root-mean-square distances
are quite small and similar to ranges of attractive pockets in their potentials.
These observations suggest that $\Delta$ and $\Omega$ are tightly bound states at this quark mass, and can be regarded qualitatively as composite states of 3 quarks rather than meson-baryon molecule states, which are already known from the phenomenological studies.

\begin{figure}[t]
    \begin{center}
        \includegraphics[width=0.49\textwidth]{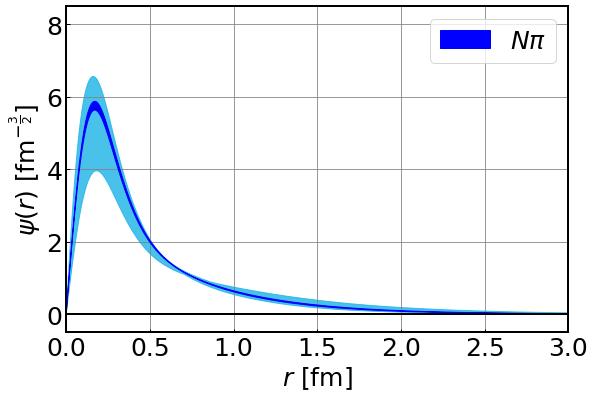}
	    \includegraphics[width=0.48\textwidth]{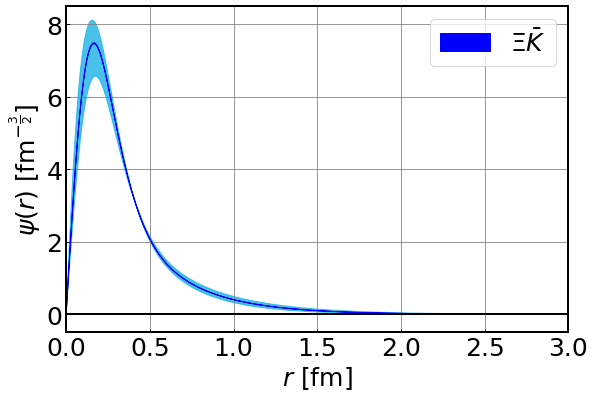}
	    \caption{Normalized wave functions for bound states in $N\pi$ (Left) and $\Xi\bar{K}$ (Right) systems.
        Dark and light blue bands represent statistical errors
          and statistical and systematic errors added in quadrature, respectively.
	    }
    \label{fig:wf}    
    \end{center}
\end{figure}

More quantitatively, however, their root-mean-square distances are larger than a range of the attractive pocket $r \approx 0.2~\textrm{fm}$
in both $N\pi$ and $\Xi\bar{K}$ potentials with the centrifugal term. (See Fig.~\ref{fig:Npi_potfit_cent}.) This can be explained by shapes of wave functions, shown in Fig.~\ref{fig:wf},
which exhibits peak structures at $r \approx 0.2~\textrm{fm}$ as well as long-range tails.

Table.~\ref{tab:part_ksq} represents the expectation value of each term in the Schr\"odinger equation Eq.~\eqref{eq:scheq_radial}.
By comparing $\Xi\bar{K}$ and $N\pi$ systems,
the kinetic term for $\Xi\bar{K}$ is positively larger than that for $N\pi$, while the potential term for $\Xi\bar{K}$ is negatively larger. 
Since the latter effect is larger than the former,
  the sum of the kinetic and potential terms ($ = \langle k^2/2\mu\rangle$) for $\Xi\bar{K}$
  is negatively larger than that for $N\pi$.
  These behaviors can be intuitively understood in the following way:
  Since the $\Xi\bar{K}$ system has a larger reduced mass,
  the system tends to be squeezed into a smaller size,
  which leads to larger effect from the attractive potential at short distances
  (but with a larger kinetic energy).
  In fact, in Fig.~\ref{fig:wf}, we observe that the $\Xi\bar{K}$ system
  has a larger peak of the wave function at short distance,
  which gives a negatively larger contribution of the potential energy,
  $\int d^3r \ \psi^{\ast}(\vb{r})V(r)\psi(\vb{r})$.

  These observations lead to an interesting picture on
  how these two systems evolve toward smaller quark masses.
  Assuming that the potentials are not so sensitive to the quark masses,
the bound state disappears and $\Delta$ becomes a resonance in the $N\pi$ system because of its small reduce mass and the broad structure of $\Delta$, while the bound state $\Omega$ remains in the $\Xi\bar{K}$ system due to its large reduced mass leading to the compact size of $\Omega$. This scenario may explain spectra observed in Nature.

\begin{table}
\begin{ruledtabular}
\begin{tabular}{c|cccc}
system   &
$\langle-\frac{1}{2\mu r}\dv[2]{r}r\rangle~\textrm{[MeV]}$&
$\langle\frac{l(l+1)}{2\mu r^2}\rangle~\textrm{[MeV]}$&
$\langle V(r)\rangle~\textrm{[MeV]}$&
$\langle\frac{k^2}{2\mu}\rangle~\textrm{[MeV]}$

 \\
\colrule
$N\pi$   & $538.1(24.3) \smqty(+136.3\\-274.8)$ 
        & $1160.3 (41.1) \smqty(+288.9\\-519.5)$
        & $-1804.1 (74.6) \smqty(+835.3\\-463.8)$
        & $-105.7 (14.3) \smqty(+61.0\\-42.2)$\\
$\Xi\bar{K}$  &  $644.0 (2.7)           \smqty (+86.6\\-130.8)$ 
        & $1207.3 (3.5) \smqty(+190.0\\-204.7)$
        & $-2078.6 (8.7) \smqty(+396.8\\-335.1)$ 
        & $-227.3 (4.2) \smqty(+66.2\\-63.3)$ 
\end{tabular}
\end{ruledtabular}
\caption{\label{tab:part_ksq}%
Expectation value of each term in Eq.~\eqref{eq:scheq_radial}.
        The first and second errors represent statistical and systematic errors, respectively.
}
\end{table}

\begin{figure}[t]
    \begin{center}
        \includegraphics[width=0.49\textwidth]{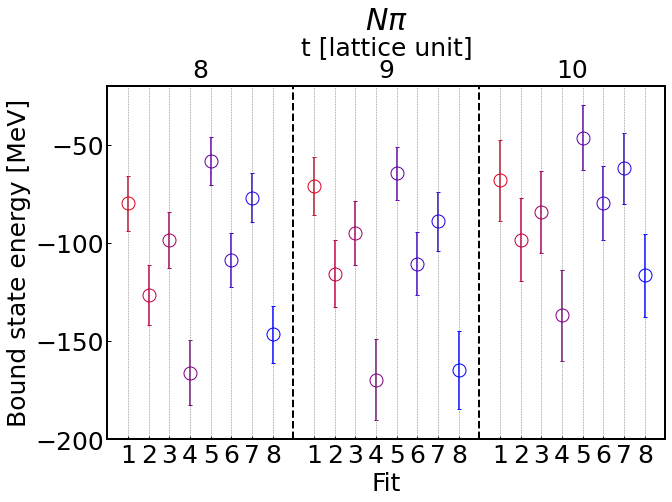}
	    \includegraphics[width=0.48\textwidth]{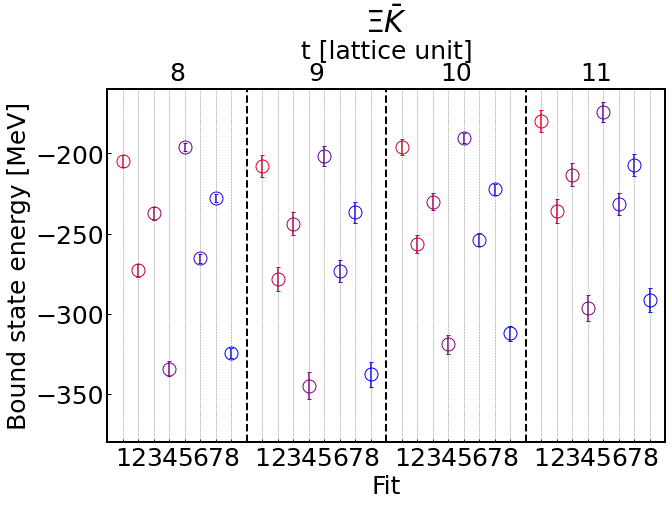}
	    \caption{Bound state energies for different fitting schemes and $t$ for $N\pi$ (Left) and $\Xi\bar{K}$ (Right). Plots are separated for different $t$ by dashed lines.
      }
	    \label{fig:be_eachcase}    
    \end{center}
\end{figure}

\begin{figure}[t]
    \begin{center}
        \includegraphics[width=0.49\textwidth]{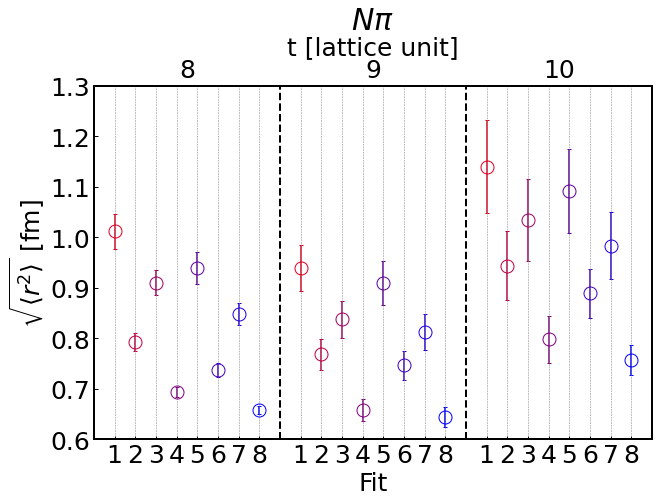}
	    \includegraphics[width=0.48\textwidth]{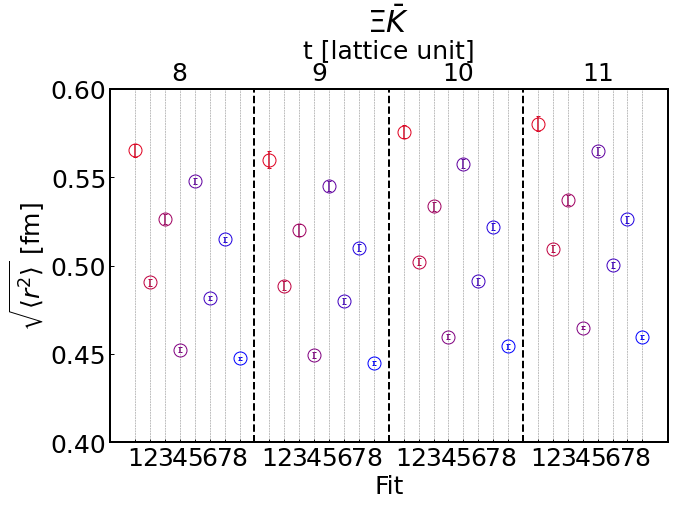}
	    \caption{Root-mean-square distances for different fitting schemes and $t$ for $N\pi$ (Left) and $\Xi\bar{K}$ (Right). Plots are separated for different $t$ by dashed lines.}
	    \label{fig:rms_eachcase}    
    \end{center}
\end{figure}

\begin{figure}[t]
    \begin{center}
        \includegraphics[width=0.7\textwidth]{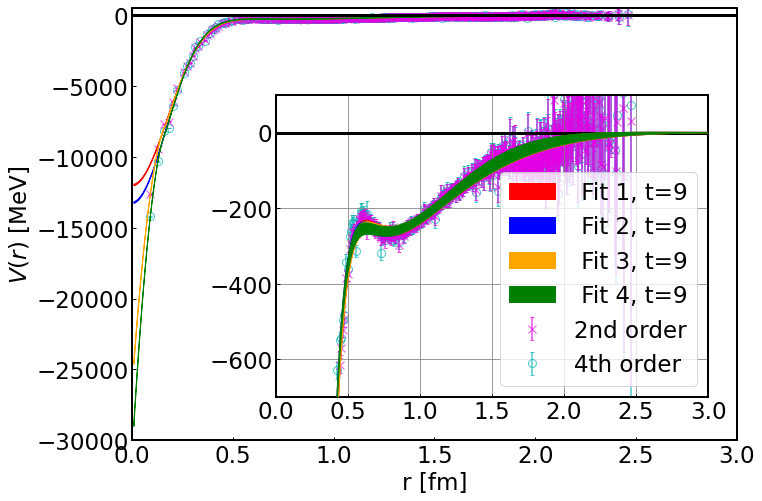}
	    \caption{Original potential data with 2nd (magenta) and 4th (cyan) order accuracies of the Laplacian term for $N\pi$ at $t=9$, and fit results for Fit 1-4 at the same $t$.}
	    \label{fig:ref_sys}    
    \end{center}
\end{figure}

Results of binding energies and root-mean-square distances suffer from quite large systematic uncertainties compared to statistical errors. Such large uncertainties come dominantly from lattice artifacts at short distances. Fig.~\ref{fig:be_eachcase} and Fig.~\ref{fig:rms_eachcase} shows binding energies and root-mean-square distances, respectively, estimated for different fitting schemes and $t$. 
A dependence on an order of the approximation for the Laplacian (2nd/4th for Fit 1/2, 3/4, 5/6, 7/8), 
or 
a dependence on a treatment of data at $r=a$ (without/with for Fit 1/3, 2/4, 5/7, 6/8), are much larger than a dependence on $t$ and a dependence on fit without/with mirrors (Fit 1/5, 2/6, 3/7, 4/8). As seen in Fig.~\ref{fig:ref_sys}, the precision of the Laplacian term affects potential data at short distances. (Compare, for example, magenta and cyan points.) Moreover, fit results without and with data at $r=a$ (without/with for Fit 1/3, 2/4) deviate from each other around the origin. Thus, large dependences of binding energies and root-mean-square distances on the accuracy of the Laplacian and the treatment of the data at $r=a$ are indeed caused dominantly by lattice artifacts at short distances, which is associated with rapid changes of potentials around the origin. On the contrary, the results are not so sensitive to finite-volume effects as seen from comparisons between Fit $i$ and Fit $i+4$ in Fig.~\ref{fig:be_eachcase} and Fig.~\ref{fig:rms_eachcase}.

\section{Conclusion}\label{sec:conclusion}
In this paper, we have studied meson-baryon interactions in P-wave $I=3/2$  $N\pi$ and $I=0$ $\Xi\bar{K}$ systems in lattice QCD using the HAL QCD method with 3-quark type source operators. 
We have employed gauge configurations at $m_{\pi}\approx 410$~MeV, where both $\Delta$ and $\Omega$ baryons exist as bound states in $N\pi$ and $\Xi\bar{K}$ systems, respectively. 
We have extracted scattering phase shifts and binding energies by solving the Schr\"{o}dinger equation with fitted potentials.

The $\Xi \bar K$ potential has a weaker attraction than the $N\pi$ while the binding energy of the $\Xi \bar K$ is larger than that of the $N\pi$, which indicates that the difference between the two interactions is not a reason for the larger binding energy of the $\Xi \bar K$, in other words, the inequality $m_{N}+m_{\pi}-m_{\Delta}<m_{\Xi}+m_{\bar K}-m_{\Omega}$. 
Instead, larger difference of the potential term between the two systems compared with the kinetic term suggests that this inequality is mainly explained by a smaller spatial size of the wave function in $\Xi\bar K$ due to a larger reduced mass, together with a strong attraction at short distances. 
This probably holds even at the physical pion mass, so that $\Delta$ exists as a resonance while $\Omega$ is a stable particle.

Root-mean-square distances of bound states in both systems are very small, which indicates that in this setup, $\Delta$ and $\Omega$ are tightly bound states and can be regarded qualitatively as 3-quark states.

Binding energies in both systems are consistent with those extracted from 2-point functions of single $\Delta$ and $\Omega$ baryons,
though systematic errors are rather large, mainly due to lattice artifacts at short distances. 
Large systematic errors are observed for the root-mean-square distances as well.
This suggests that further improvement is desirable for the HAL QCD method to analyze states which behave like a single particle rather than a bound state of two hadrons, such as $\Delta$ and $\Omega$ in this study.
This is because a short distant part of the HAL QCD potentials between two hadrons, which is relevant for such a compact state,
suffers severely from lattice artifacts,
while  a direct extraction of a single hadron mass from a 2-point correlation function of a single hadron operator
is rather insensitive to such problems.  
In a setup where $\Delta$ appears as a resonance, on the other hand, the HAL QCD method is expected to become more efficient.

In this work, we have performed the LO analysis in the derivative expansion of non-local potentials in the HAL QCD method. 
In the present study, it is implied that the LO analysis is sufficient since
binding energies from potentials are consistent with those extracted from 2-point correlation
functions within systematic uncertainties.
However, when the systematic uncertainties from lattice discretization are improved,
the next-leading order (NLO) analysis might be required as well,
since the states are deeply bound below the threshold in this lattice setup.
NLO analysis may be also necessary to study the $\Delta$ as a resonance in future, since a resonance peak appears much higher energy than the threshold. The study in this direction is presented in Ref.~\cite{Akahoshi:2021sxc}.

\section{Acknowledgements}
We thank the PACS-CS Collaboration~\cite{PACS-CS:2008bkb} and ILDG/JLDG~\cite{Amagasa:2015zwb} for providing us their gauge configurations. We use lattice QCD code of Bridge++~\cite{Ueda:2014rya, bridge++url} and its optimized version for the Oakforest-PACS by Dr. I. Kanamori~\cite{Kanamori:2018hwh}. Our numerical calculation has been performed on HOKUSAI BigWaterfall at RIKEN and Oakforest-PACS in Joint Center for Advanced HighPerformance Computing (JCAHPC). This work is supported in part by the HPCI System Research Project (Project ID: hp200108(FY2020), hp210061(FY2021)), Multidisciplinary Cooperative Research Program in CCS, University of Tsukuba, the Grant-in-Aid of the MEXT for Scientific Research (Nos. JP16H03978, JP18H05236, JP18H05407, JP19K03879), ``Program for Promoting Researches on the Supercomputer Fugaku'' (Simulation for basic science: from fundamental laws of particles to creation of nuclei), and Joint Institute for Computational Fundamental Science (JICFuS).
Y.A. is supported in part by the Japan Society for the Promotion of Science (JSPS). K.M. is supported in part by JST SPRING, Grant Number JPMJSP2110, and by the Japan Society for the Promotion of Science (JSPS). 
S.A. is supported in part by he Grant-in-Aid of the MEXT for Scientific Research (Nos. JP16H03978, JP18H05236).
We thank other members of the HAL QCD Collaboration for fruitful discussions. We also appreciate Dr. J. Bulava for providing us several meson-baryon operators projected onto the irreps of the cubic group, which are used for checking our code in this work.
 
\appendix

\section{Singular behaviors of the $N\pi$ potential and the sink smearing}\label{sec:ss}
\subsection{$N\pi$ potential with point sink quarks}
\begin{figure}[t]
    \begin{center}
        \includegraphics[width=0.49\textwidth]{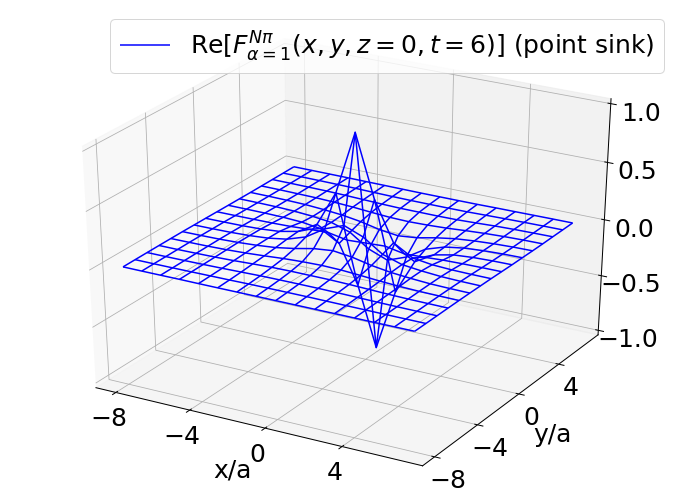}
	    \includegraphics[width=0.48\textwidth]{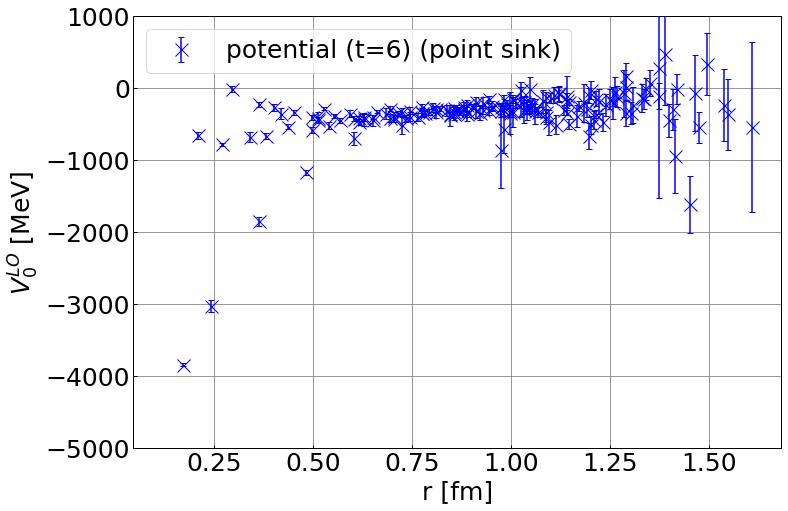}
	    \caption{(Left) Three-dimensional plot of $x/a$, $y/a$, and a real part of the $N\pi$ 3-point function Eq.~\eqref{eq:def_of_Npi3pt} at $t=6$, $z=0$, and $\alpha = \uparrow$ for point sink quark operators. 
        It is normalized such that the maximum value is $1$. 
        (Right) The corresponding leading-order potential.}
	    \label{fig:Npi_point}    
    \end{center}
\end{figure}

We have sometimes observed singular behaviors of a potential between hadrons in the HAL QCD method
if two hadrons involve quark-antiquark creation/annihilation diagrams at the valence level.
For example, 
Fig.~\ref{fig:Npi_point} (Left) shows a real part of the normalized 3-point function in the $N\pi$ system for point sink quark operators in our trial calculation using $2+1$-flavor configurations by the CP-PACS and JLQCD Collaborations~\cite{JLQCD:2007xff} at $a\approx0.12~\textrm{fm}$
and  $m_{\pi} \approx 870~\textrm{MeV}$ on a $16^3\times 32$  lattice, while Fig.~\ref{fig:Npi_point} (Right) is the corresponding leading-order potential. Due to a complicated structure with rapid changes of the 3-point function around the origin, 
the corresponding potential shows multi-valued behavior at short distances,
which is, however, NOT caused by contaminations of higher partial waves due to a cubic box (long distance effects),
as discussed below.
Due to quark-antiquark annihilations possible in this channel, 
the operator-product expansion~\cite{Aoki:2009pi,Aoki:2010kx,Aoki:2010uz,Aoki:2013zj} predicts a behavior of the 3-point function at $r \to 0$ as
 \begin{eqnarray}\label{eq:Yoverrcube}
 F(\vb{r}) \propto \frac{1}{r^3}Y_{l,m}(\Omega).
\end{eqnarray}
$Y_{l,m}(\Omega)$ is the spherical harmonics with the angular momentum of the system, and thus $F(\vb{r})$ has non-trivial angular dependence except for $l=0$. We find that the behavior of Eq.~\eqref{eq:Yoverrcube} for $l=1$ at short distances looks very similar to a plot in Fig.\ref{fig:Npi_point} (Left),
and the corresponding potential also becomes multi-valued as in Fig.\ref{fig:Npi_point} (Right),
if the number of discrete data at short distances is too small to reproduce a smooth single-valued behavior.
Thus the problem is caused by short distance discretization effects. Note that a similar singular behavior was already observed in the $I=1$ P-wave $\pi\pi$ system~\cite{Akahoshi:2021sxc}. 

\begin{figure}[t]
    \begin{center}
        \includegraphics[width=0.49\textwidth]{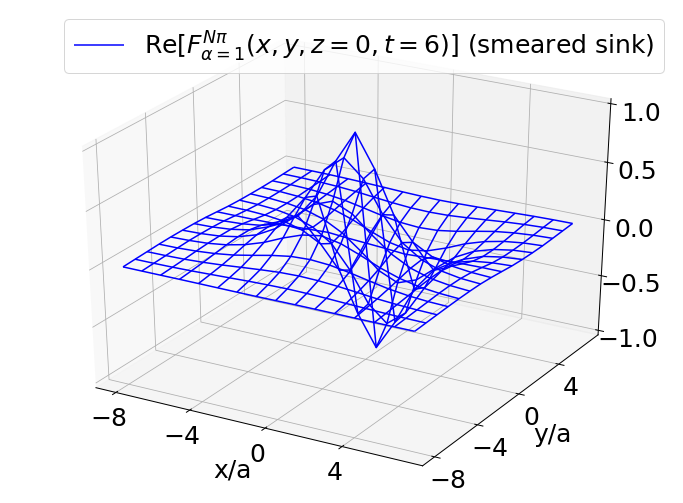}
	    \includegraphics[width=0.48\textwidth]{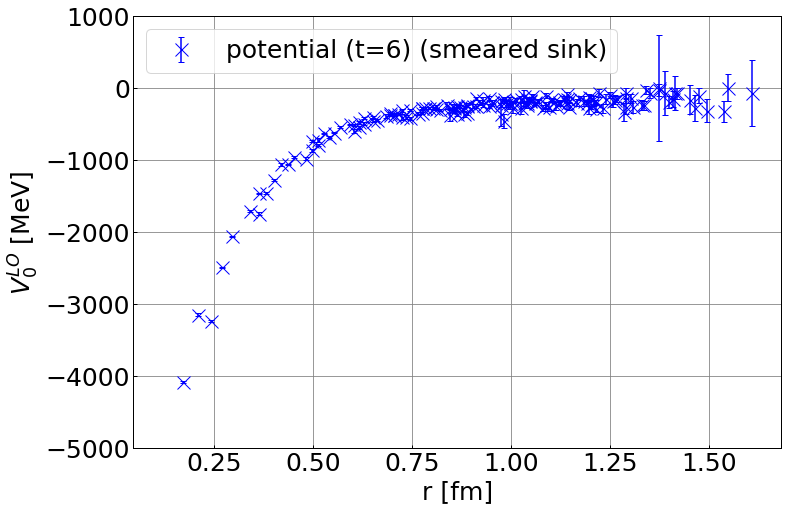}
	    \caption{ Same as Fig.~\ref{fig:Npi_point}  but for smeared sink quark operators. }
	    \label{fig:Npi_smear}    
    \end{center}
\end{figure}

One of possible prescriptions to tame singular behaviors at short distances is the sink smearing at the quark level, introduced in ~\cite{Akahoshi:2021sxc}. 
Fig.\ref{fig:Npi_smear} (Left) represents the 3-point function with smeared sink quark operator, while Fig.\ref{fig:Npi_smear} (Right) is 
the corresponding potential. As expected, sink smearings
make the behavior of the wave function at short distances much smoother, so that the potential becomes almost single-valued even at short distances where discrete data points are sparsely located.

\subsection{Non-locality in potentials with smeared sink operators}
Although smeared sink operators suppress singular behaviors in potentials at short distances, 
they may enhance non-locality in potentials. This may cause large systematic errors in the leading order analysis, 
as was observed in the case of the LapH method~\cite{HALQCD:2017xsa},

In order to quantify effects of non-locality caused by our sink smearing,  
we compare scattering phase shifts between point and smeared sink operators 
in the same setup as the main text, taking $I=1$ S-wave $\Xi\bar{K}$ system.

\begin{figure}[t]
    \begin{center}
        \includegraphics[width=0.7\textwidth]{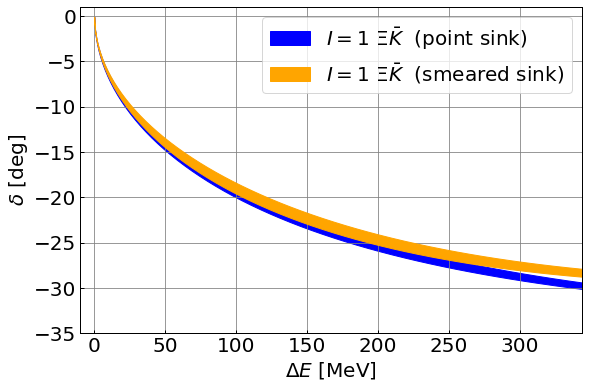}
	\caption{Phase shifts in $I=1$ $\Xi\bar{K}$ system calculated form potentials at $t=8$. Blue (orange) bands show results with point (smeared) sink operators.
        }
	    \label{fig:sstest}    
    \end{center}
\end{figure}

Comparisons in Fig.~\ref{fig:sstest} show that phase shifts from different sink operators agree with each other 
at $\Delta E \lesssim 100$ MeV, where $\Delta E$ is an energy difference from the threshold.
This suggests that non-locality due to the sink smearing
has negligible effects on physical observables in the low energy region in this setup.

\section{Exact relativistic formula in the time-dependent HAL QCD method with unequal masses}\label{sec:exactrelhal}

In the main text, we employ Eq.~(\ref{eq:scheq_rela_approx}) to obtain the potential, 
in which the relativistic effect is expressed as the Taylor series in terms of time derivatives.
While only up to 2nd order derivatives are required if the masses of two hadrons are the same~\cite{Ishii:2012ssm},
the series run up to an infinite order if two hadrons have unequal masses.
In this appendix, we present an alternative formula in which relativistic effect can be exactly included
with at most the 3rd time derivatives.

\subsection{Naive formula}
First, we directly apply Eq.(\ref{eq:ksqDeltaW}) to the time-dependent HAL QCD method~\cite{Ishii:2016zsf}. Multiplying Eq.(\ref{eq:scheqofelastic}) by $(\Delta W_{n}/M + 1)^2$, we have 
\begin{dmath}
 \left(\mathcal{P}(\Delta W_{n}) - \left(1+\frac{\Delta W_{n}}{M}\right)^2 H_{0} \right) A_{n} \Psi^{W_{n}}_{\alpha}(\vb{r})e^{-\Delta  W_{n}t} 
 =\int d^3r' \ U_{\alpha\beta}(\vb{r},\vb{r}')\left(1+\frac{\Delta W_{n}}{M}\right)^2 A_{n} \Psi^{W_{n}}_{\beta}(\vb{r'})e^{-\Delta  W_{n}t}.
\end{dmath}
Rewriting $\Delta W_{n}$ in terms of the time derivative, we obtain
\begin{dmath}
 \left(  \mathcal{P}\left(-\pdv{t}\right) - \left(1-\frac{1}{M}\pdv{t}\right)^2 H_{0} \right) A_{n} \Psi^{W_{n}}_{\alpha}(\vb{r})e^{-\Delta  W_{n}t} 
 =\int d^3r' \ U_{\alpha\beta}(\vb{r},\vb{r}')\left(1-\frac{1}{M}\pdv{t}\right)^2 A_{n} \Psi^{W_{n}}_{\beta}(\vb{r'})e^{-\Delta  W_{n}t}.
\end{dmath}
Taking summation over $n$, we reach the exact relativistic form of the relation as
\begin{eqnarray}
 \left(\mathcal{P}\left(-\pdv{t}\right) - \left(1-\frac{1}{M}\pdv{t}\right)^2 H_{0} \right) R_{\alpha}(\vb{r},t) 
 =\int d^3r' \ U_{\alpha\beta}(\vb{r},\vb{r}')\left(1-\frac{1}{M}\pdv{t}\right)^2 R_{\beta}(\vb{r}',t),
\end{eqnarray}
where
\begin{eqnarray}
\mathcal{P}\left(-\pdv{t}\right)
= -\pdv{t} + \frac{\mu+M}{2\mu M}\pdv[2]{t} - \frac{1}{2\mu M}\pdv[3]{t} + \frac{1}{8\mu M^2}\pdv[4]{t}.
\end{eqnarray}

If we use the derivative expansion for the non-locality of the potential, the leading-order potential in this formulation reads
\begin{eqnarray}\label{eq:LOpotential_rel_gen}
V^{\textrm{LO}}(r) =
\frac{\left(\mathcal{P}\left(-\pdv{t}\right) - \left(1-\frac{1}{M}\pdv{t}\right)^2 H_{0} \right) R_{\alpha}(\vb{r},t)}{\left(1-\frac{1}{M}\pdv{t}\right)^2 R_{\alpha}(\vb{r},t)},
\end{eqnarray}
which requires at most 4th order time derivative terms.

\subsection{Improved formula}
There is a sophisticated technique for a reduction of the highest order in the time derivatives. We decompose $\mathcal{P}\left(-\pdv{t}\right)$ into a term proportional to $(1-\frac{1}{M}\pdv{t})^2$ and a remnant as
\begin{eqnarray}
\begin{aligned}
&2\mu \mathcal{P}\left(-\pdv{t}\right) 
= 
\left(1-\frac{1}{M}\pdv{t}\right)^2 2\mu\left[-\pdv{t} + b\pdv[2]{t}\right] \\
&+ 2\mu\left[ \left(-b+\frac{1}{2\mu }-\frac{3}{2M}\right)\pdv[2]{t}
-\frac{1}{M}\left(-2b+\frac{1}{2\mu }-\frac{1}{M}\right)\pdv[3]{t}
+\frac{1}{M^2}\left(-b+\frac{1}{8\mu }\right)\pdv[4]{t}
\right],   
\end{aligned}
\end{eqnarray}
where we introduce an arbitrary parameter $b$,
such that terms proportional to $b$ are summed up to zero and the above identity formula holds for any $b$.
Taking $b=\frac{1}{2M}(1+c \,\delta^2)$ with another arbitrary parameter $c$, where $\delta=(m_M-m_B)/M$, we obtain
\begin{dmath}\label{eq:exact_rel_w_beta}
\left(1-\frac{1}{M}\pdv{t}\right)^2 \left[-\nabla^2 R_{\alpha}(\vb{r},t)+2\mu \int d^3r' \  U_{\alpha\beta}(\vb{r},\vb{r}') R_{\beta}(\vb{r}',t) \right] 
=\left(1-\frac{1}{M}\pdv{t}\right)^2 2\mu \left[-\pdv{t} + \left( \frac{1}{2M}+\frac{\delta^2}{2M}c+\frac{\delta^4}{8\mu}c\right)\pdv[2]{t}\right]R_{\alpha}(\vb{r},t)
+ \delta^2\left[ \left(-\frac{1}{4}c+1\right)\pdv[2]{t}
-\frac{1}{M}\left(-\frac{1}{2}c+1\right)\pdv[3]{t}
+\frac{1}{M^2}\left(-\frac{1}{4}c+\frac{1}{4}\right)\pdv[4]{t}\right]R_{\alpha}(\vb{r},t).
\end{dmath}

We then decompose the potential as 
\begin{eqnarray}\label{eq:exact_rel_w_beta_sum}
U_{\alpha\beta}(\vb{r},\vb{r}') =  U^{(0)}_{\alpha\beta}(\vb{r},\vb{r}')+ U^{(1)}_{\alpha\beta}(\vb{r},\vb{r}'),
\end{eqnarray}
where $U^{(i)}_{\alpha\beta}(\vb{r},\vb{r}')$ ($i=0,1$) are defined by
\begin{eqnarray}
&&\left[-\nabla^2 R_{\alpha}(\vb{r},t)+2\mu \int d^3r' \  U^{(0)}_{\alpha\beta}(\vb{r},\vb{r}') R_{\beta}(\vb{r}',t) \right] \nonumber \\
&=& 2\mu \left[-\pdv{t} + \left( \frac{1}{2M}+\frac{\delta^2}{2M}c+\frac{\delta^4}{8\mu}c \right)\pdv[2]{t}\right]R_{\alpha}(\vb{r},t), 
\end{eqnarray}
\begin{dmath}\label{eq:exact_rel_w_beta_remnant}
\left(1-\frac{1}{M}\pdv{t}\right)^2 2\mu \int d^3r' \ U^{(1)}_{\alpha\beta}(\vb{r},\vb{r}') R_{\beta}(\vb{r}',t) \\
=\delta^2\left[ \left(-\frac{1}{4}c+1\right)\pdv[2]{t}
-\frac{1}{M}\left(-\frac{1}{2}c+1\right)\pdv[3]{t}
+\frac{1}{M^2}\left(-\frac{1}{4}c+\frac{1}{4}\right)\pdv[4]{t}\right]R_{\alpha}(\vb{r},t) .
\end{dmath}

If we use the derivative expansion for the non-locality of the potential, the leading-order potential in this decomposition  reads
\begin{eqnarray}\label{eq:LO_exact_rel_w_beta_sum}
 V^{LO}(\vb{r}) =  V^{(0)LO}(\vb{r})+ V^{(1)LO}(\vb{r}),
\end{eqnarray}
where
\begin{eqnarray}\label{eq:LO_exact_rel_w_beta_eachterm}
V^{(0)LO}(\vb{r})
&=&\frac{1}{R_{\alpha}(\vb{r},t)}
\left[\left(-\pdv{t} + \left( \frac{1}{2M}+\frac{\delta^2}{2M}c+\frac{\delta^4}{8\mu}c\right)\pdv[2]{t}\right)-H_{0} \right]R_{\alpha}(\vb{r},t),\\
V^{(1)LO}(\vb{r})
&=&\frac{\frac{\delta^2}{2\mu}
\left[ \left(-\frac{1}{4}c+1\right)\pdv[2]{t}
-\frac{1}{M}\left(-\frac{1}{2}c+1\right)\pdv[3]{t}
+\frac{1}{M^2}\left(-\frac{1}{4}c+\frac{1}{4}\right)\pdv[4]{t}\right]R_{\alpha}(\vb{r},t)}
{\left(1-\frac{1}{M}\pdv{t}\right)^2 R_{\alpha}(\vb{r},t)}.
\label{eq:exact_rel_w_beta_remnant_LO}
\end{eqnarray}

If we set $c=1$, the 4th order time derivative term in Eq.~(\ref{eq:exact_rel_w_beta_remnant}) 
or Eq.~(\ref{eq:exact_rel_w_beta_remnant_LO}) vanishes, so that we need only  3rd order terms at most.

\subsection{Comparison of potentials among semi-relativistic formula and exact relativistic formulae}

\begin{figure}[t]
    \begin{center}
        \includegraphics[width=0.49\textwidth]{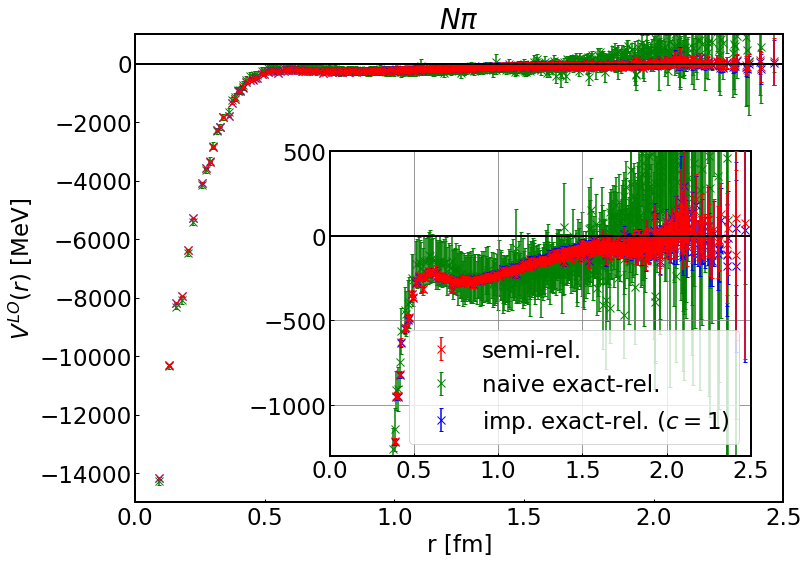}
	    \includegraphics[width=0.48\textwidth]{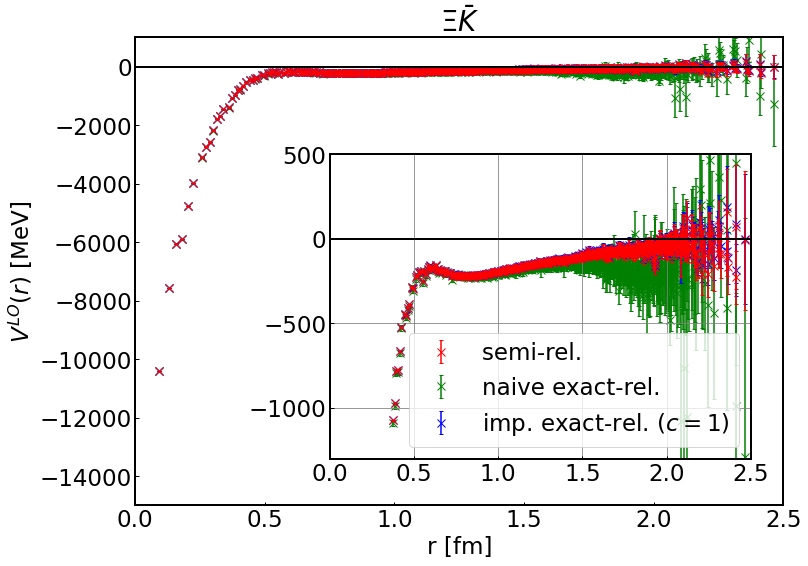}
	    \caption{The $N\pi$ potential at $t=9$ (Left) and the $\Xi\bar{K}$ potential at $t=10$ (Right) estimated from Eq.(\ref{eq:LOpotential_dec}) 
        (Red), and Eq.(\ref{eq:LOpotential_rel_gen}) (Green) and Eq.(\ref{eq:LO_exact_rel_w_beta_sum}). (Blue). 
        Note that blue points are almost the same as red points and thus are almost invisible in the figure.
      }
	    \label{fig:semirelvsexactrel}    
    \end{center}
\end{figure}

Fig.~\ref{fig:semirelvsexactrel} shows a comparison of the $N\pi$ and $\Xi\bar{K}$ potentials estimated from Eq.~(\ref{eq:LOpotential_dec}), and those from Eq.~(\ref{eq:LOpotential_rel_gen}) and Eq.~(\ref{eq:LO_exact_rel_w_beta_sum}).
In the calculation of Eq.~(\ref{eq:LOpotential_dec}), 4th and higher order time derivative terms are truncated.
All three results are consistent with each other within statistical errors, though results from Eq.~(\ref{eq:LOpotential_rel_gen}) has larger statistical errors, which are caused by statistical fluctuations of the 4th order time derivative terms appeared in the numerator. 
Since results from Eq.~(\ref{eq:LOpotential_dec}) and Eq.~(\ref{eq:LO_exact_rel_w_beta_sum}) agree with each other 
within comparable errors, systematic uncertainties for the truncation in Eq.~(\ref{eq:LOpotential_dec}) are well under control.

\nocite{*}
 
\bibliography{Npipreprint_v3}

\end{document}